\newif\ifjournal
\journaltrue

\ifjournal
  \documentclass{aa}
  \usepackage{graphicx,times}
\else
  \documentclass{paper}
\fi

\newcommand{\kpc}{h^{-1}\,\mathrm{kpc}}
\newcommand{\muG}{\mu\mathrm{G}}
\newcommand{\V}[1]{\vec{#1}}

\begin{document}


\title{Evolution and structure of magnetic fields in simulated galaxy
clusters}
\ifjournal
  \author{Klaus Dolag\inst{1} \and Matthias Bartelmann\inst{1} \and
    Harald Lesch\inst{2}}
  \offprints{K.~Dolag}
  \institute{Max-Planck-Institut f\"ur Astrophysik, P.O.~Box 1317,
    D--85741 Garching, Germany \and
    Universit\"ats-Sternwarte M\"unchen, Scheinerstr.~1, D--81679
    M\"unchen, Germany}
  \date{}
\else
  \author{Klaus Dolag$^1$, Matthias Bartelmann$^1$, and Harald
    Lesch$^2$\\
    $^1$Max-Planck-Institut f\"ur Astrophysik, P.O.~Box 1317, D--85741
    Garching, Germany\\
    $^2$Universit\"ats-Sternwarte M\"unchen, Scheinerstr.~1, D--81679
    M\"unchen, Germany}
  \date{}
\fi

\newcommand{\abstext}
  {We use cosmological magneto-hydrodynamic simulations to study the
   evolution of magnetic fields in galaxy clusters in two different
   cosmological models, a standard-CDM and a $\Lambda$-CDM model. We
   show that the magnetic field strength profiles closely follow the
   cluster density profiles outside a core region of radius
   $\sim200\kpc$. The magnetic field has a correlation length of order
   $50\kpc$ and reverses on scales of $\sim100\kpc$ along typical
   lines-of-sight. The power spectrum of the magnetic field can well
   be approximated by a steep power law with an exponent of
   $\sim-2.7$. The mean magnetic field in the cluster cores grows
   roughly exponentially with decreasing redshift,
   $B\sim10^{-2.5z}\,\muG$. Merger events have a pronounced effect on
   magnetic field evolution, which is strongly reflected in measurable
   quantities like the Faraday rotation. The field evolution in the
   two different cosmologies proceeds virtually identically. All our
   cluster models very well reproduce observed Faraday rotation
   measurements when starting with nG seed fields.}

\ifjournal
  \abstract{\abstext
  }
\else
  \begin{abstract}\abstext\end{abstract}
\fi

\maketitle

\section{Introduction}

While Faraday-rotation measurements clearly show that galaxy clusters
contain magnetic fields which are smooth on scales of order $100\kpc$,
their origin and their evolution are largely unclear. Fields of that
scale must be related to the formation and evolution of the host
cluster rather than the kinematics and internal evolution of the
cluster galaxies. It is therefore an interesting question in what way
galaxy clusters deform, compress and amplify the seed fields
pervading the material that finally ends up in cluster cores, what
magnetic field configurations are prevalent in final, virialised
clusters, and whether the cosmological background model has any
significant effect on the evolution of intracluster fields.

We address these questions with cosmological, magnetohydrodynamic
cluster simulations. The numerical code used for this study uses
smooth-particle hydrodynamics (SPH) for the magnetised intracluster
gas and purely gravitational interaction between the particles of the
dominant dark-matter component. The technical and numerical properties
of our ``Grape-MSPH'' code, its performance in various tests, and
results of a first application to cluster evolution were described in
an earlier paper (Dolag et al. 1999). Here, we extend this earlier study in several
ways. First, we perform cluster simulations in two different
cosmological models, a standard-CDM (SCDM) model with $\Omega=1$ and
$\Lambda=0$, and a $\Lambda$CDM model with $\Omega=0.3$ and
$\Lambda=0.7$, to explore the influence of the cosmological background
model. Second, we focus on the structure of the intracluster magnetic
fields, quantified by radial field-strength profiles, correlation and
field-reversal lengths, and the magnetic-field power spectrum. Third,
we investigate how magnetic fields grow during cluster evolution, and
how they are affected by cluster mergers. Finally, we compare the
Faraday rotation predicted by our cluster models with the largest
available sample of Faraday-rotation measurements.

\section{Numerical technique and initial conditions}

\subsection{GrapeMSPH}

Our cosmological MHD code ``Grape-MSPH'' was described in detail in
\cite{DBL99}. Therefore, we only give a brief summary here. The code combines
the merely gravitational interaction of a dominant dark-matter
component with the hydrodynamics of a magnetised gaseous
component. The gravitational interaction of the particles is evaluated
on GRAPE boards (cf.~Sugimoto et al.~1990), while the gas dynamics is
computed in the SPH approximation (Lucy 1977; Monaghan 1992). The
original ``GrapeSPH'' code (Steinmetz 1996) was augmented by an
implementation of the (ideal) magneto-hydrodynamic equations
describing the evolution of the magnetic fields carried by the
gas. Throughout, we assume that the electric conductivity of the gas
is infinite, which implies that the magnetic fields are frozen into
the gas flow. The back-reaction of the magnetic fields on the gas via
the Lorentz force is fully included. The numerical viscosity required
by SPH to properly capture shocks is chosen such that angular-momentum
transport in presence of shear flows is carefully controlled. The SPH
kernel width is automatically adapted to the local number density of
SPH particles, which results in an adaptive spatial resolution of the
code.

Extensive tests of the code were performed and described in
\cite{DBL99}. The code succeeds in solving the co-planar MHD Riemann
problem posed by Brio \& Wu (1988). Although the simulated magnetic
field is not strictly divergence-free, monitoring shows that
$\nabla\cdot\vec B$ always remains negligible compared to the magnetic
field divided by a typical length scale of $\vec B$ (see \cite{DBL99}
for more detail). The code also assumes the intracluster medium to be
an ideal gas with an adiabatic index of $\gamma=5/3$. Although
foreseen and implemented in the code, we neglect cooling here. One
reason is that we cannot avoid the cooling catastrophe without a
substantial source of heat. Most of the gas cools and ends up in an
unphysically condensed central lump. Moreover, the cool gas fraction
strongly depends on the resolution of the simulation. In addition, a
substantial amount of numerical heating is expected due to our limited
mass resolution.

The surroundings of the clusters are important because their
gravitational tidal fields affect the overall cluster structure and
the merger history of the clusters during formation. Therefore, the
cluster simulation volumes are surrounded by a layer of boundary
particles whose purpose it is to accurately represent the tidal fields
of the cluster neighbourhood.

\begin{table}[t]
\begin{center}
\begin{tabular}{|c|cccccc|}
\hline
 model & $H_0$ & $\Omega_\mathrm{m}^0$ & $\Omega_\Lambda^0$ &
 $\sigma_8$ & $f_\mathrm{bary}$ & Source\\
\hline
 SCDM & 0.5 & 1.0 & 0.0 & 1.2 & 5\% & (1)\\
 $\Lambda$CDM & 0.7 & 0.3 & 0.7 & 1.05 & 10\% & (2)\\
\hline
\end{tabular}
\end{center}
\caption{Parameters of the SCDM and the $\Lambda$CDM model. The SCDM
  clusters are COBE normalised, which is in conflict with the local
  cluster abundance. However, this does not affect our specific
  results because we are studying the magnetic field growth in
  individual clusters. Initial conditions are kindly provided by: (1)
  Bartelmann \& Steinmetz 1996, (2) Eke et al.~1998.
\label{tab:models}}
\end{table}

\subsection{Initial conditions}

We set up initial conditions for two CDM-dominated model universes
(standard CDM, SCDM; and flat, low-density CDM, $\Lambda$CDM), whose
parameters are listed in Table~\ref{tab:models}. For the SCDM models,
the central region of $\sim70\,{\rm Mpc}$ (comoving) are filled with
dark-matter particles with a mass of $3.2\times10^{11}\,M_\odot$ each,
mixed with an equal number of gas particles whose mass is twenty times
smaller. The central region is surrounded by collisionless boundary
particles whose mass increases outward to mimic the tidal forces of
the neighbouring large-scale structure. Including the region filled
with boundary particles, the simulation volume is a sphere with a
(comoving) diameter of $\sim300\,{\rm Mpc}$. For the $\Lambda$CDM
models, the dark matter particles have masses of
$8.1\times10^{10}\,M_\odot$ each, while the mass of the gas particles
is ten times smaller. Including the region filled with boundary
particles, the simulation volume in this case has a comoving diameter
of $\sim600\,{\rm Mpc}$. The simulations labelled as ``double'' are
taken from the SCDM models, but have twice the number of particles
with half the original mass. For each cosmology, we
used ten different realisations of the density-fluctuation field at an
initial redshift $z_\mathrm{ini}=15$ (SCDM, kindly provided by
M.~Steinmetz) and $z_\mathrm{ini}=20$ ($\Lambda$CDM, kindly provided
by J.~F.~Navarro). Note that the growth of fluctuations is slowed down
in a $\Lambda$CDM model compared to the SCDM model. The different
initial redshifts compensate for that, so that the initial
fluctuations in linear expectation grow by the same amount in both
cosmologies.

Each of these realisations evolves such that it contains clusters of
different final masses and in different dynamical states at the final
redshift $z=0$. We simulate each of these clusters with up to five
different initial configurations of the magnetic field, which are
listed in Table.~2. In addition, one set (SCDM) is computed with its
mass resolution increased by a factor of two to test for resolution
effects in our results. In total, we simulate and study 100 cluster
models.

\begin{table}[t]
\begin{center}
\begin{tabular}{|c|c|c|c|}
\hline
 model & $B_\mathrm{ini}$
 & $\langle B_\mathrm{final}\rangle_\mathrm{core}^\mathrm{SCDM}$
 & $\langle B_\mathrm{final}\rangle_\mathrm{core}^{\Lambda\mathrm{CDM}}$\\
\hline
low     & $0.2\times10^{-9}\,{\rm G}$ & $0.4\;\mu{\rm G}$ & $0.3\;\mu{\rm G}$\\
chaotic & $0.2\times10^{-9}\,{\rm G}$ & $0.4\;\mu{\rm G}$ & ---\\
double  & $0.2\times10^{-9}\,{\rm G}$ & $1.0\;\mu{\rm G}$ & ---\\
medium  & $1.0\times10^{-9}\,{\rm G}$ & $1.1\;\mu{\rm G}$ & $0.8\;\mu{\rm G}$\\
high    & $5.0\times10^{-9}\,{\rm G}$ & $2.5\;\mu{\rm G}$ & $2.0\;\mu{\rm G}$\\
\hline
\end{tabular}
\end{center}
\caption{Initial magnetic fields (Col. 2) and the resulting final
  mean magnetic field strengths in the cluster cores (Cols. 3 and 4
  for SCDM and $\Lambda$CDM models, respectively). The final values
  are an average over the gas particles central region (enclosed by a
  radius of $0.1\,r_{\mathrm vir}$) across the sample of ten clusters
  simulated for each cosmology. Note that in the case of doubled mass
  resolution, the amplification of the initial field is significantly
  increased because shear flows are better resolved. This means that
  the initial field required to reproduce the observations with our
  simulations are only upper limits. Apart from that, there is no
  indication that the increased resolution changes the characteristics
  of the final magnetic field.\label{tab:bfield}}
\end{table}

Lacking any detailed knowledge on the origin of primordial magnetic
seed fields, we explore two extreme cases of initial field
configurations. In one case (``{\em homogeneous\/}''), we assume that
the field is initially constant throughout the cluster volume. In the
other case (``{\em chaotic\/}''), we let the initial field orientation
vary randomly from place to place, subject only to the condition that
$\nabla\cdot\vec B=0$. The initial field strengths in both cases are
determined by setting the mean field energy densities. To study the
effects of changing the mean initial energy density of the magnetic
seed field, we ran simulations using three choices for this quantity,
labelled ``{\em low\/}'', ``{\em medium\/}'' and ``{\em high\/}''.

To gain some insight into performance of the numerics, we also ran a
set of simulations with doubled mass resolution, labelled ``{\em
double\/}''. The magnetic field configurations of these runs are the
same as in the ``{\em homogeneous\/}'' case. Of course, this is not
meant to replace a full resolution study. Here, we are limited by the
capacity of our GrapeMSPH code, because the Grape design restricts the
number of gas particles. We also ran a set of control simulations for
each cosmological model without magnetic field, labelled ``{\em
no\/}''. In total, we have a set of ten realisations for each
cosmology, simulated with six different magnetic field setups in the
SCDM cosmology, and with four different magnetic field setups in the
$\Lambda$CDM cosmology, leading to the 100 simulated clusters
mentioned before.

Table~\ref{tab:bfield} summarises the initial field set-ups and the
mean field strengths in the cluster cores. The initial field strengths
are of order $10^{-9}\,$G, the final field strengths of micro-Gauss
order are averaged over the gas particles within spheres of
$0.1\,r_{\mathrm vir}$ and across the sample of ten clusters. For some
aspects of the analysis of the simulations to be described later on,
we also use ten less massive objects identified close to the main
clusters. As they are relatively more poorly resolved in the
simulations, the results from this smaller objects have to be treated
with caution.

\begin{figure*}[p]
\begin{center}
 \includegraphics[width=0.8\textwidth]{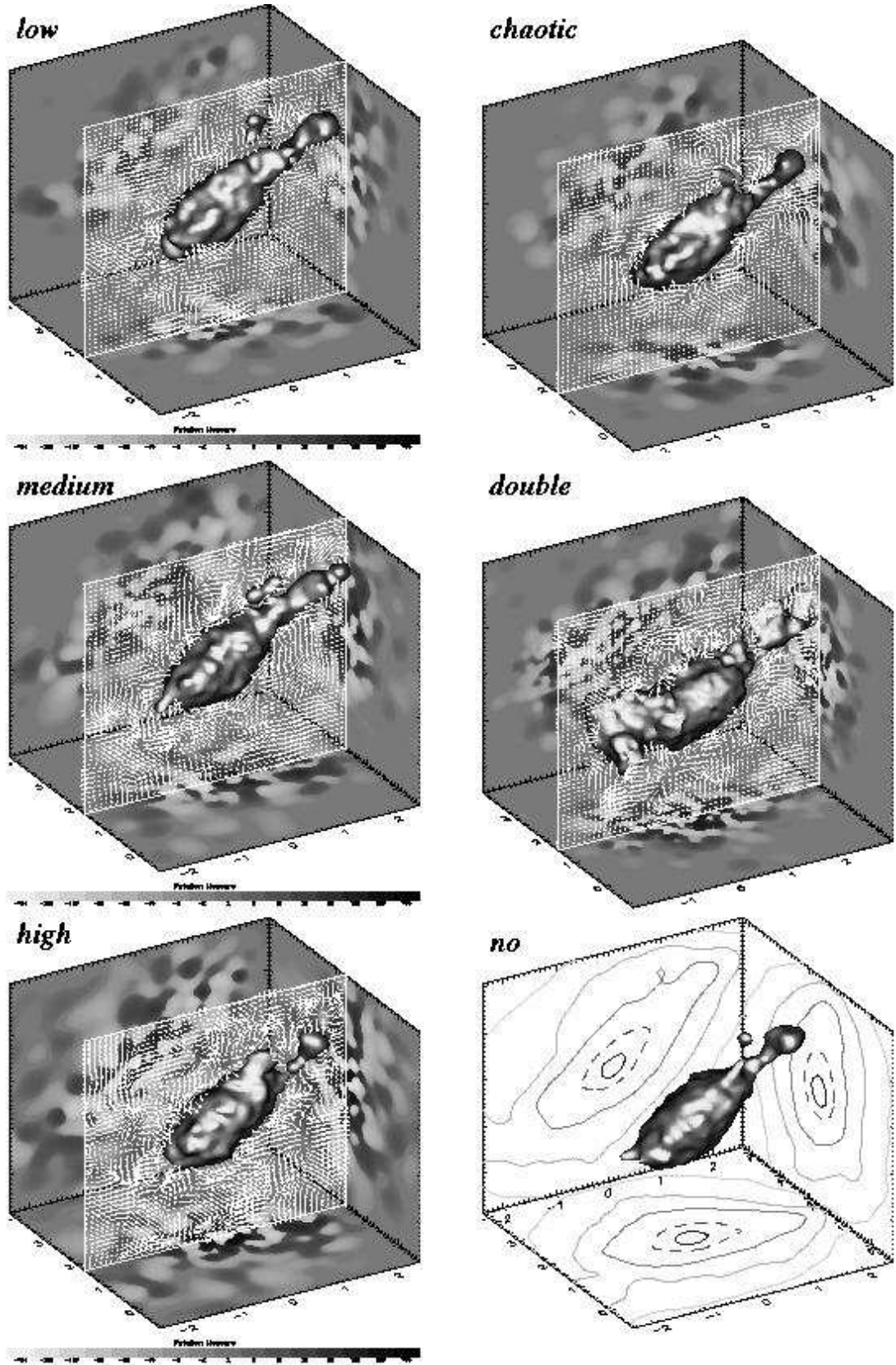}
\end{center}
\caption{The figure shows versions of the same SCDM galaxy cluster for
  six different initial magnetic field models as indicated in the
  plot, including one version without magnetic field for
  reference. Plotted are the projected rotation measure maps, an
  iso-density surface and a cut through the magnetic field.}
\label{fig:rm}
\end{figure*}

Furthermore, we point out that as we are including the back-reaction
of the magnetic field on the gas, different initial field strengths
and configurations are changes to the initial conditions from which
the clusters develop, hence their evolution proceeds in a somewhat
different way. This implies that even a low magnetic field can alter
the dynamical time scale of a simulated cluster to some degree.
Therefore, when comparing simulated clusters with different initial
magnetic fields at equal redshifts, one has to keep in mind that they
can be in slightly different evolutionary stages according to their
individual dynamical time scale. Of course this has an even more
pronounced effect on the models with doubled mass resolution. In this
case, it is not the magnetic field but the addition of new particles
at intermediate positions which changes the initial conditions and
leads to slightly different clusters which evolve in a slightly
different fashion. Figure~\ref{fig:rm} provides a visual impression of
the differences between six different simulations of one cluster in a
pre-merger phase, starting from the same SCDM cluster initial
conditions. The simulations use four different configurations of the
initial magnetic field, one run without magnetic field, and one with
doubled mass resolution.

\begin{table}[t]
\begin{center}
\begin{tabular}{|c|c|c|c|c|}
\hline
 SCDM  & \multicolumn{2}{c|}{$r<0.1r_\mathrm{vir}$} & 
         \multicolumn{2}{c|}{$r<1.0r_\mathrm{vir}$} \\
\cline{2-5}
       & A [$\mu G$]& $\alpha$ & A [$\mu G$]& $\alpha$ \\
\hline

low     & $5.9\times10^{-4}$ & 2.5 & $9.4\times10^{-4}$ & 2.2 \\
chaotic & $4.0\times10^{-4}$ & 2.6 & $8.6\times10^{-4}$ & 2.3 \\
double  & $2.5\times10^{-3}$ & 2.6 & $1.0\times10^{-3}$ & 2.6 \\
medium  & $5.0\times10^{-3}$ & 2.1 & $6.1\times10^{-3}$ & 1.9 \\
high    & $3.1\times10^{-2}$ & 1.7 & $3.6\times10^{-2}$ & 1.4 \\

\hline
\hline
 $\Lambda$CDM & \multicolumn{2}{c|}{$r<0.1r_\mathrm{vir}$} & 
                \multicolumn{2}{c|}{$r<1.0r_\mathrm{vir}$} \\
\cline{2-5}
       & A [$\mu G$]& $\alpha$ & A [$\mu G$]& $\alpha$ \\
\hline

low     & $1.1\times10^{-4}$ & 3.0 & $3.0\times10^{-4}$ & 2.5\\
medium  & $1.1\times10^{-3}$ & 2.5 & $2.5\times10^{-3}$ & 2.0\\
high    & $1.0\times10^{-2}$ & 2.1 & $1.5\times10^{-2}$ & 1.7\\

\hline
\end{tabular}
\end{center}
\caption{Parameters (amplitude $A$ and slope $\alpha$) of power laws
  best fitting the relation between final mean magnetic field strength
  and mean temperature in the simulated clusters. Numbers for
  different cosmologies, different magnetic field setups and different
  radii for the averaging volume are determined. The different slopes
  fitted to the different models are all consistent within the model
  uncertainties.\label{tab:b_t}}
\end{table}

\section{\label{sec:4} Structure of the final magnetic field}

As reported earlier (Dolag et al.~1999), the process of cluster
formation and the surrounding large scale structure entirely determine
the magnetic field properties in the simulated clusters. As shown
before, all information contained in the structure of the seed field
is completely wiped out in the process of the collapse. This means
that the final magnetic field properties obtained reflect our
understanding of the formation of galaxy clusters, with the underlying
general assumption that the magnetic field in galaxy clusters results
from the amplification of weak seed fields.

\subsection{Amplification by shear flows}

Apart from the compression of the magnetic field frozen into the gas,
the magnetic field is also substantially amplified by shear flows as
the gas is accreted by the cluster. A necessary condition for this
amplification mechanism is that the velocity difference across the
boundary layer exceeds the local Alfv\'en speed, which is proportional
to the field strength, so that the condition is more easily satisfied
for weaker fields. This means that the relative field amplification
becomes smaller as the fields grow. As the shear flows are largely
driven by gradients in the dark matter potential, this mechanism is
quite independent of pressure support by the magnetic field and
therefore reduces the relative amplification much earlier than local
equipartition is reached.

\begin{table}[t]
\begin{center}
\begin{tabular}{|c|c|c|c|c|}
\hline
model & $\langle l_\mathrm{final}\rangle_\mathrm{core}^\mathrm{SCDM}
[\mathrm{kpc}]$ & RMS \\
\hline
low B   & 62.6 &  9.8 \\
chaotic & 60.8 & 22.0 \\
double  & 55.5 & 16.5 \\
medium B& 51.3 & 12.9 \\
high B  & 46.1 & 18.2 \\
\hline
\end{tabular}
\end{center}
\caption{Cols. 2 and 3 show typical field-reversal scales in kpc
  (physical units) for the magnetic fields and their {\em rms\/}
  deviation for the clusters in the SCDM model. The values are
  calculated using all particles in cluster-centric spheres with one
  tenth of the viral radius, and averaged across the sample of ten
  clusters per cosmological model.}
\label{tab:lscale}
\end{table}

\subsection{Final field strength}

\begin{figure*}[ht]
 \includegraphics[width=0.49\textwidth]{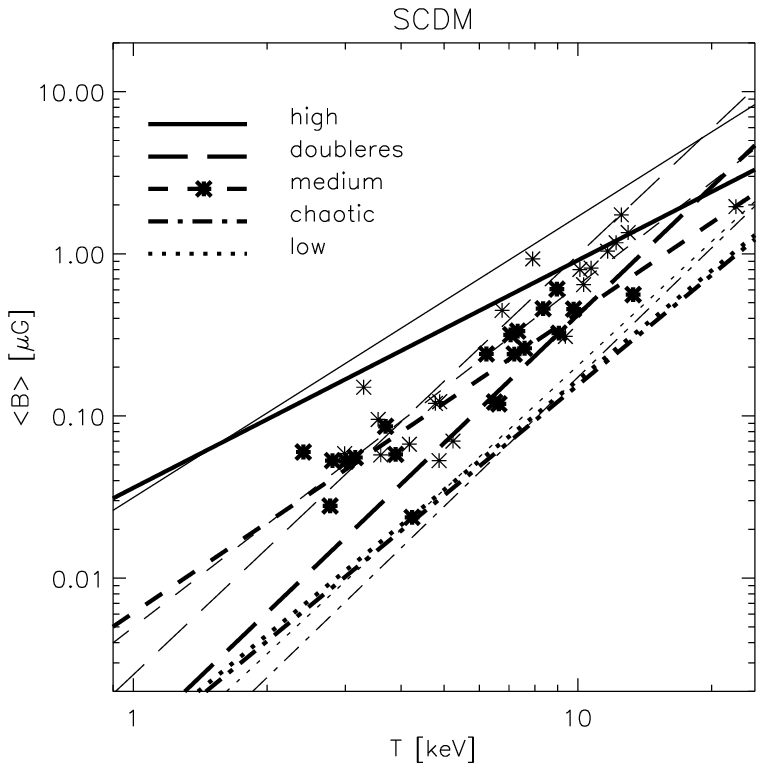}
 \includegraphics[width=0.49\textwidth]{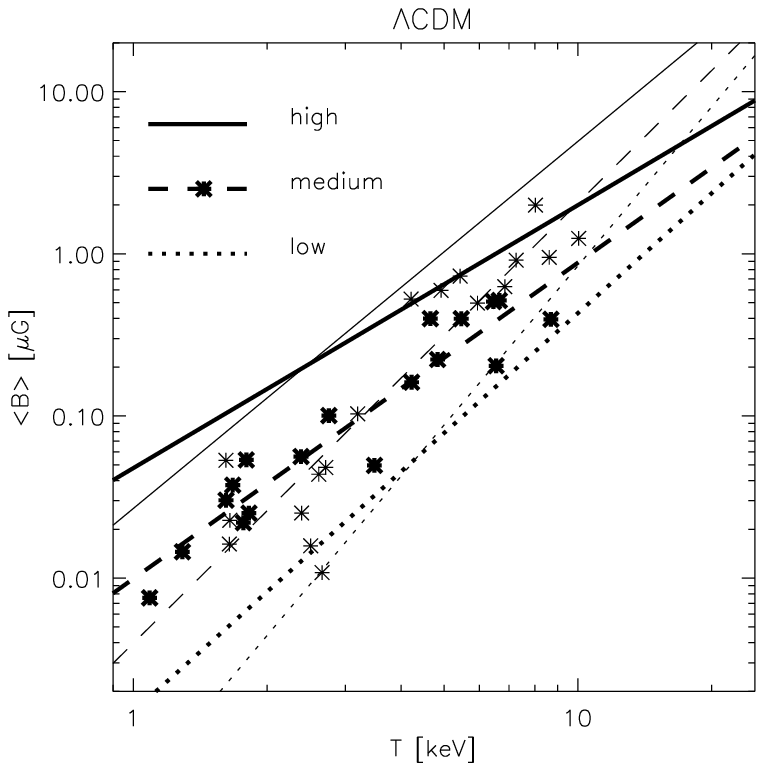}
\caption{The mean magnetic field is shown as function of the mean mass
  weighted temperature inside the clusters. Here, the ten less massive
  objects identified in the simulations are included in the
  analysis. The symbols belong to the ``{\em medium\/}'' model, the
  lines represent the power laws fitted to all models for the
  different magnetic seed fields. Both quantities are calculated
  within 10 per cent (thin symbols and lines) and 100 per cent (thick
  symbols and lines) of the virial radius. The left side represents
  the SCDM models, the right side the LCDM models, respectively.}
\label{fig:b_t}
\end{figure*}

One quantity of primary theoretical interest is the final magnetic
field strength in galaxy clusters predicted by our models. Of course,
it depends on the strengths of the seed fields assumed in the
simulations. The final magnetic field strengths also depend on the
final cluster mass and on the dynamical state of the galaxy cluster
because mergers are known to amplify the magnetic field (Roettiger
et al.~1999, Dolag et al.~1999).

The simulations also predict that the magnetic field decreases with
increasing cluster-centric distance. Therefore, we compute the mean
magnetic field strengths within 10 per cent and 100 per cent of the
virial radius of the final cluster. Figure~\ref{fig:b_t} shows how the
mean magnetic field strength scales with temperature for three sets of
initial field strengths in our SCDM simulations. In order to increase
the temperature range of our cluster set, we also include ten less
massive objects found near the most massive clusters in our
simulations. We thus ended up with a total of twenty different objects
for every initial magnetic field configuration in both cosmologies.

We fitted this relation in all sets of simulations by a power law,
$\langle B\rangle=A\,T^\alpha$. The values of $A$ and $\alpha$ we
obtained for all our simulations in both cosmologies are summarised in
Table~\ref{tab:b_t}. In Fig.~\ref{fig:b_t}, the different symbols and
line types distinguish between the different initial field models. The
thin and thick symbols and lines show results calculated within 10 per
cent and 100 per cent of the virial radius, respectively.

As Table~\ref{tab:b_t} shows, this relation tends to flatten when the
magnetic field strength is averaged within a larger volume, and also
when the initial seed magnetic field strength is increased. These
trends are not significant given the uncertainties of the models, but
they may have the same reason as the flattening of the magnetic field
profiles towards the cluster cores, which will be discussed below. It
is simply the fact that smaller magnetic fields can more easily be
amplified by shear flows than already grown magnetic fields, as
discussed before. These relations again show that there is no
difference in the final states of the simulation sets with ``{\em
chaotic\/}´´ and ``{\em homogeneous\/}´´ initial field configurations,
which confirms that the information on the initial field configuration
gets entirely wiped out during the process of cluster collapse. There
is also no visible difference between the two cosmologies. Doubling
the mass resolution 
does not significantly change the final field structure either,
although the field amplification is higher.

\begin{figure*}[ht]
 \includegraphics[width=0.49\textwidth]{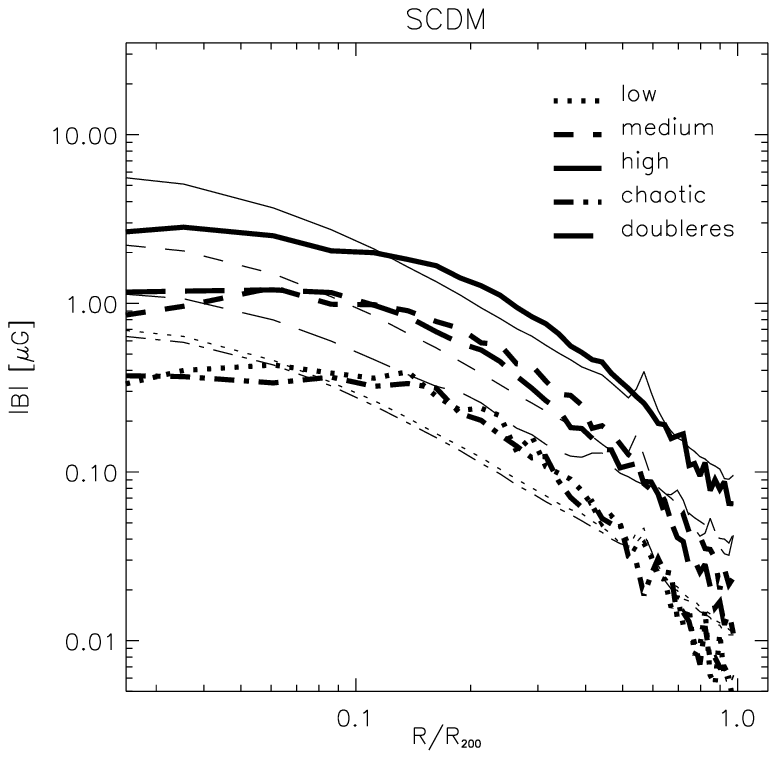}
 \includegraphics[width=0.49\textwidth]{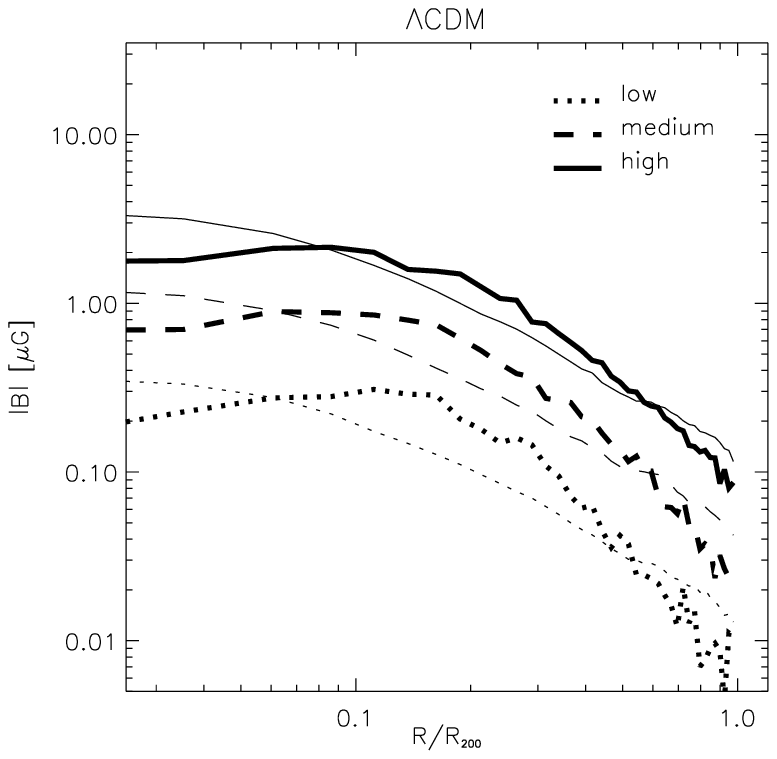}
\caption{The two panels show the radial profiles of the magnetic field
  strengths (thick lines) in the simulated, final clusters. The
  profiles are obtained by averaging across the sample of ten
  clusters. The averaged gas density profiles, raised to the power of
  $2/3$, are over-plotted for comparison as the thin lines, arbitrarily
  normalised to the amplitude of the magnetic-field profiles at the
  virial radius. The left panel shows the SCDM simulations for the
  various initial field configurations, as labelled in the plot. The
  right panel shows the results for the $\Lambda$CDM models.}
\label{b_prof_all}
\end{figure*}

Another interesting result of the simulations is the relation between
the magnetic field strength and the gas density in the clusters.
Figure~\ref{b_prof_all} shows the mean magnetic field strength in
spherical shells centred on the cluster centre as function of
cluster-centric distance in units of the virial radius. Plotted are the
profiles at redshift zero, but the shape does not change significantly
between redshift one and zero. The gas density profiles, averaged
across the sample of ten clusters and raised to the power of $2/3$,
are over-plotted in arbitrary units to match the magnetic field
strength amplitude at the virial radius (dashed lines).

The average magnetic field strength profiles in the outer regions of
the clusters (beyond $\sim0.3\,r_\mathrm{vir}$) follow the density
profile very well, while they flatten off towards the cluster core,
where the magnetic field profiles are shallower than the density
profiles. Plotting the density profile raised to the power of $2/3$
demonstrates that in the outer parts of the cluster, the average
magnetic field strength is more strongly amplified as expected from
compression alone ($\propto \rho^{2/3}$), thus shear flows have to
play an important role there. In the centre, the field amplification
is reduced. This is due to the fact that the higher magnetic field
strength in the cluster centres slows down the further field
amplification compared to the outer cluster regions. There are two
reasons for that. First, the motions of the gas are less coordinated
in the cluster cores compared to the large-scale motion of the
infalling gas in the outer parts, so that amplification by ordered
shear flows is suppressed. Moreover, increasingly important
back-reaction of the field on the gas flow counteracts further
amplification. Second, the magnetic field is also substantially
amplified by shear flows as the gas is accreted by the cluster. A
necessary condition for such an amplification is that the relative
velocity on both sides of the boundary layer exceeds the local
Alfv\'en speed which is proportional to the field strength, so that
the condition is more easily satisfied for the weaker fields embedded
in the infalling material in the outer regions of the clusters than
for the stronger fields near cluster centres.

This is also responsible for flattening the relation between
temperature and magnetic field strengths for increasing initial
magnetic field strength. Table~\ref{tab:bfield} lists the final
magnetic field strengths in the cores of the simulated cluster sample
for the different initial magnetic field configurations. Again, the
chaotic and the homogeneous initial field configurations lead to the
same radial magnetic field strength profiles. As observed before in
Dolag et al.~(1999), the final magnetic field structure is entirely
driven by the collapse of the galaxy cluster, and any information
about the initial field structure is completely erased. This justifies
the use of homogeneous initial conditions for the magnetic field
setup, as they are numerically easier to handle. The profiles do not
change significantly when the particle number is doubled.

These results on the radial magnetic field strength profiles and the
dependence of mean field strengths on the cluster temperature may help
to reconcile the observations in different galaxy clusters, which seem
to give somewhat different values for the magnetic field for different
clusters observed. In addition, the difference between magnetic field
inferred from Faraday rotation measurements on the one hand (testing
the line-of-sight average) and the combination of radio emission with
hard X-ray observations on the other hand (measuring volume averaged
magnetic fields) could be bridged using our simulations, although the
relativistic electron density needs to be assumed. Comparing the mean
magnetic field within the virial radius and one tenth of the virial
radius, the data points in Fig.~\ref{fig:b_t} show that these values
differ already by factors of five, or more.

\subsection{Field reversals}

\begin{figure}[t]
 \includegraphics[angle=0,width=0.5\textwidth]{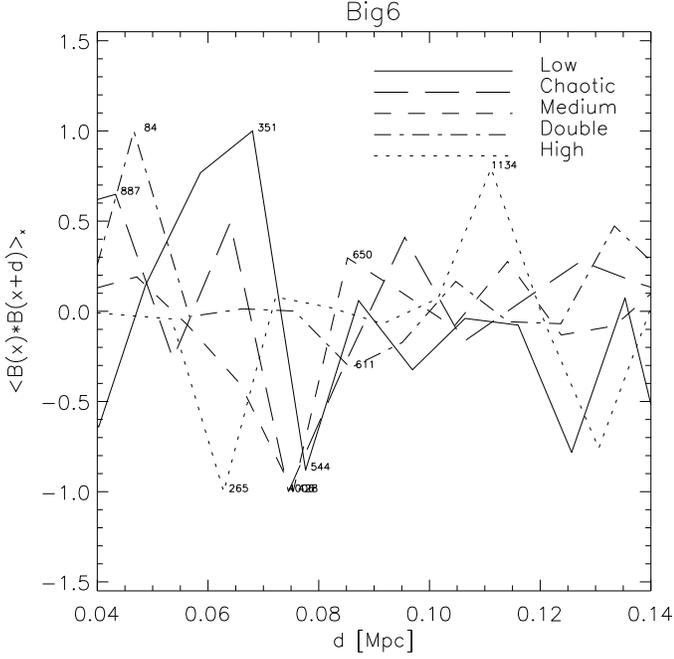}
\caption{The figure shows the autocorrelation function of the magnetic
  field, $\langle\V{B}(\V{x})\cdot\V{B}(\V{x}+\V{d})\rangle$ as a
  function of $|\V{d}|$ for one SCDM cluster, simulated with five
  different initial conditions for the magnetic field, as indicated in
  the plot. The numbers attached to the minima and maxima indicate the
  number of particle pairs used to compute the mean value. Only
  particles within ten per cent of the virial radius are taken into
  account. All curves are normalised to fall within $\pm1$ within the
  region shown. Table~\ref{tab:lscale} lists the mean inferred
  autocorrelation length scale across our cluster sample and its {\em
  rms\/} deviation.}
\label{fig_corr}
\end{figure}

The formation of large scale structure in the surroundings of a
cluster also imprint some structure on the magnetic field of galaxy
clusters. Large scale flows can order magnetic fields parallel with
the flow. Also the field compression in shock fronts orders the
magnetic fields to some degree. Since the shocks compress the medium
in only one direction, the magnetic fields in the shocked gas tend to
lie within the surface of the shock shell. On the other hand, the more
chaotic motion of the gas inside cluster cores tends to destroy the
order imposed on the magnetic field during cluster collapse. We thus
expect a complicated behaviour of the field structure and a range of
autocorrelation length scales. It is of interest for the
interpretation of rotation measures on what scales field orientations
reverse inside cluster cores. Our simulations predict how magnetic
fields can be ordered in the process of large scale structure
formation and gravitational collapse. In addition, there might be
smaller scale contributions from the outflows and the cocoons of the
radio galaxies, from which Faraday rotations are measured.

In order to infer this typical length scale in the simulated clusters, we
cut out the central sphere with one tenth of the virial radius
of each cluster and calculate the autocorrelation function
\begin{equation}
  \xi(\V{d})=\langle\V{B}(\V{x})\cdot\V{B}(\V{x}+\V{d})\rangle
\end{equation}
by identifying all possible pairs of particles, sorting them by their
separation, and binning them. The mean is then calculated over the
number of particles in each distance bin. This function should be
positive for scales smaller than that of typical field reversals, then
go through a minimum at this typical length scale, and then damp due
to the mixing of different length scales. This mixing, the presence of
shocks, and the layers between shear flows, already affect this
measurement at the length scale of interest, which render the
determination (and even the meaning) of the length scale dubious. On
larger scales, it gets confused by the mixture of different length
scales present in the cluster. Therefore, the minimum correlation
length scale is difficult to measure.

Figure~\ref{fig_corr} shows the autocorrelation function for one
simulated SCDM cluster. At small scales, this function shows the
positive value expected, but it is already very noisy. Note that the
innermost distance bin is narrower than the average inter-particle
distance, and are therefore likely to be in shocks.

Although the magnetic field in the shocks is preferentially aligned
tangentially to its surface, the orientation within this surface can
be very tangled. Accordingly, the innermost bin is very noisy and can
even yield negative values, as seen in the ``{\em low\/}'' model. At
somewhat larger distances, the autocorrelation function becomes
positive, as expected. Then, at almost twice this distance,
$\xi(\V{d})$ becomes negative, also as expected. Clearly, the field
autocorrelation function is determined by the structure of the cluster
and shows a very similar behaviour for different initial magnetic field
configurations in that it consistently reaches a pronounced minimum
between 60 and 80 kpc. The numbers attached to the individual lines
denote the number of particles used to calculate the mean within this
distance bin. At the first minimum, several hundred particles are used
to compute the SPH average. The autocorrelation function reaches zero
only beyond the plotted range.

\begin{figure}[t]
  \includegraphics[angle=0,width=0.5\textwidth]{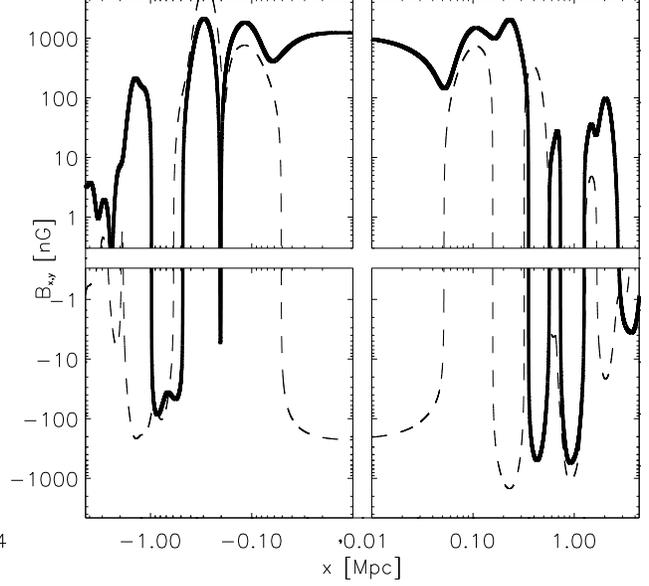}
\caption{This figure shows the magnetic field component in the $x$
  (solid line) and $y$ directions (dashed line) along a line-of-sight
  through one of the simulated clusters. This line was chosen parallel
  to the $x$ axis and passing through the cluster centre. The envelope
  of the curves follows the general radial decrease in the magnetic
  field strength. It is clearly seen that field reversals are
  relatively more frequent in the cluster centre, and relatively less
  frequent away from the cluster core.}
\label{fig_bsight}
\end{figure}

Since the autocorrelation signal is the result of mixing structures
with different length scales (due to substructure), regions simulated
at varying spatial resolutions (because of the adaptive SPH kernel),
and volumes with varying magnetic field strength (due to the field
structure in the cluster), it is not at all straightforward to read
off a characteristic autocorrelation scale from such plots for all
clusters. In order to obtain a more robust estimator, we tried to read
off the first two extrema and the zero, and constructed the
appropriately weighted mean
\begin{equation}
  l=(2\,l_\mathrm{drop}+l_{\mathrm{zero}_1}+
     2/3\,l_\mathrm{min}+1/2\,l_{\mathrm{zero}_2})/4
\end{equation}
of the positions of the extrema to get the length scale $l$ rather
than trying to identify the location of the first minimum
only. Interpreting the signal like a wave and taking the distance
between the zero crossing as the length-scale of the magnetic field,
$l_\mathrm{drop}$ in this equation is the length scale where the
signal starts to decrease, which is an indicator of half the typical
length of the magnetic field $l_{\mathrm{zero}_1}$, where the signal
is expected to drop to zero. The next minimum $l_\mathrm{min}$ should
reflect $2/3$ of the magnetic field reversal scale. We may also get a
second minimum, $l_{\mathrm{zero}_2}$, reflecting twice the magnetic
field reversal scale, before the signal fades out due to the reasons
mentioned above.

Table~\ref{tab:lscale} summarises the field reversal scales found in
the simulations. The table lists the averages across the sample of ten
clusters for each set of simulations and their {\em rms\/}
deviations. Clearly, the reversal scale does not depend on the
structure and the strength of the initial magnetic field. Doubling the
particle resolution does not yield significantly different answers
either.

\subsection{Field structure}

The length scale on which the magnetic field typically reverses its
direction is important for the interpretation of Faraday rotation
measurements. Since the Faraday rotation measure is an integral of the
magnetic field component parallel to the line-of-sight, weighted by
the electron density, it is also important to know how the field
reversal scale changes across the cluster. Figure~\ref{fig_bsight}
shows one line of sight passing through the centre of one of the
simulated clusters. The solid and dashed lines show the $x$ component
(along the line-of-sight) and $y$ component (perpendicular to the
line-of-sight) of the magnetic field along the line-of-sight,
respectively. The line-of-sight was chosen to go through the centre of
mass of the cluster and follows the $x$ axis. Notice that the plot is
composed of four panels, where the regions near zero are cut out to
allow a logarithmic scale.

\begin{figure}[t]
  \includegraphics[angle=0,width=0.5\textwidth]{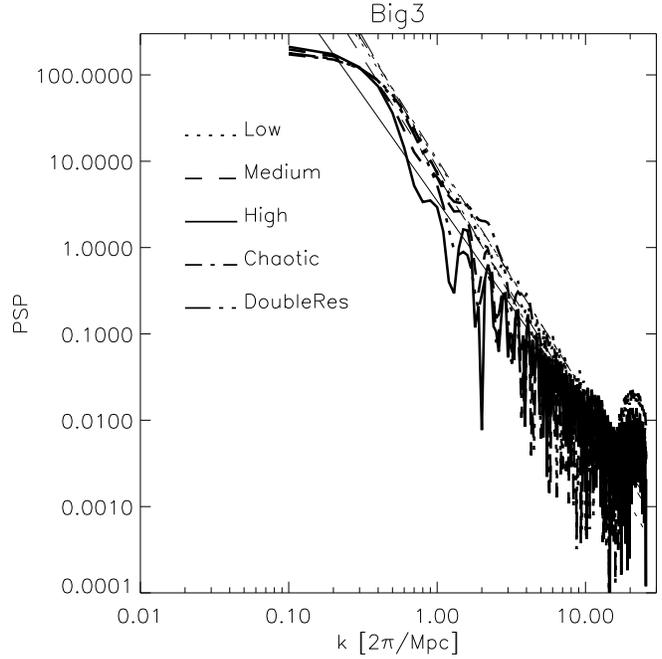}
\caption{This figure shows the magnetic-field power spectra calculated
  from the simulation of one galaxy cluster for different initial
  magnetic field configurations, distinguished by different line types
  as indicated in the plot. For each simulation, the best-fitting
  power law is overlaid.}
\label{fig_psp}
\end{figure}

It can be read off the figure how frequently the magnetic field
changes it direction, and how the reversal length scale changes with
the distance from the cluster centre. The envelope of the curves in
this plot nicely follows the radial decline of the magnetic field
strength in our simulations. It is also clearly seen that the magnetic
field reverses relatively less often in the outer parts of the
cluster. While we count $23$ extrema for the $x$ component of the
magnetic field within $\pm4\,\mbox{Mpc}$ (in physical units), there
are already 15 extrema within $\pm1\,\mbox{Mpc}$ (in physical units).

This effective increase in the reversal length scale has to be taken
into account equally well as the decrease of the magnetic field
strength with radius when inferring magnetic fields from Faraday
rotation measurements. These effects tend to reduce the magnetic field
strengths inferred from current observations.

In order to improve our understanding of the statistical properties of
the field structure, we calculated the power spectra of the magnetic
field in the simulated galaxy clusters. To this end, we calculated the
auto-correlation function of the field strength within a cube of
$10\,\mbox{Mpc}$ comoving side length, centred on the cluster centre,
using an equidistant grid of $80^3$ cells on which all quantities are
calculated by means of the SPH formalism.

\begin{table}[t]
\begin{center}
\begin{tabular}{|c|c|c|c|c|}
\hline
 model & $a$ & $b$ & $\langle b\rangle$ & RMS$_b$ \\
\hline
low     &  9.9   &  -2.7 &  ---       & --- \\
chaotic & 10.6   &  -2.7 &  ---       & --- \\
double  &  8.5   &  -3.0 &  ---       & --- \\
medium  &  6.8   &  -2.7 &  -2.7      & 0.2 \\
high    &  3.5   &  -2.4 &  ---       & --- \\
\hline
\end{tabular}
\end{center}
\caption{Best-fitting parameters for power laws, $a\,k^b$, to the
  magnetic-field power spectra. The table summarises the values for
  different initial magnetic field configurations in the SCDM models.}
\label{tab:pspindex}
\end{table}

In this type of simulation, the dynamical range of spatial scales is
not high enough to fully resolve the correlation function over a
satisfactory range of length scales. Nonetheless, we could compute
the correlation function and the power spectra over one and a half
orders of magnitude in length scale, within which it nicely follows a
power law. The hump towards small $k$ values (i.e.~large radii) can
qualitatively be understood. First, the correlation function rises
towards the typical distance between major objects within the
simulation, e.g.~the cluster and the next major object falling onto
it. In addition, the correlation function in a cubical simulation
volume is over-predicted for scales approaching half the cube side
length (i.e.~the Nyquist length) because one particle of each pair is
most probably positioned near the cluster centre. On even larger
scales, one particle of each pair most likely falls outside the
cluster core, and therefore the correlation function drops rapidly. On
small scales, the resolution of the cube used, and the resolution of
the simulation itself, restrict the calculation of the correlation
function and thus also the computation of the power spectra.

Nonetheless, it is possible to calculate the power spectra over a
reasonably wide range in wave number, and compare the results for
different magnetic field configurations. Figure~\ref{fig_psp} shows
the power spectrum calculated from one of the simulated clusters for
different initial magnetic field configurations. The power spectra
look quite the same for all models.

The exponents of the best-fitting power laws for one set of
simulations range from $-2.3$ to $-3.1$, with a mean of $-2.7$. This
value is surprisingly low. For Kolmogorov turbulence, $-5/3$ would be
expected. A fully established MHD turbulence would produce $-3/2$, and
even in the extreme case of a young, incompletely established MHD
turbulence, we would expect $-2$ (cf.~\cite{EH84} and references
therein).

There are two cases discussed in the literature which result in a
smaller power-law slope. First, for a 2-dimensional Navier-Stokes
turbulence (\cite{Bi93}), an exponent of $-3$ would be expected. As
shear flows play a major role in amplifying the magnetic field, it is
very likely that we see our cluster simulations in a state, where the
turbulence induced by shear flows evolves into a Kolmogorov
turbulence. The second case, which leads to a power-law index of $-4$
(\cite{EH84}), is turbulence in a very viscous medium. SPH is known to
be viscous, so this may be a possibility to explain the small exponent
found, but as the artificial viscosity in SPH is constructed to vanish
in shear flows, this is unlikely to be the complete
explanation. To clarify this issue, detailed tests remain to be
done.
Table~\ref{tab:pspindex} summarises the power-law fits to
the power spectra in the different models. In particular, the spectra
for the chaotic and the equivalent homogeneous initial field
configurations are indistinguishable. Also, the power spectra for the
run with doubled mass resolution do not differ, which argues against a
dominant impact of the SPH viscosity on the behaviour of the magnetic
field in the simulations. There is also no visible change in the power
laws with changes in the magnetic field strength.

\begin{figure}[t]
  \includegraphics[angle=0,width=0.5\textwidth]{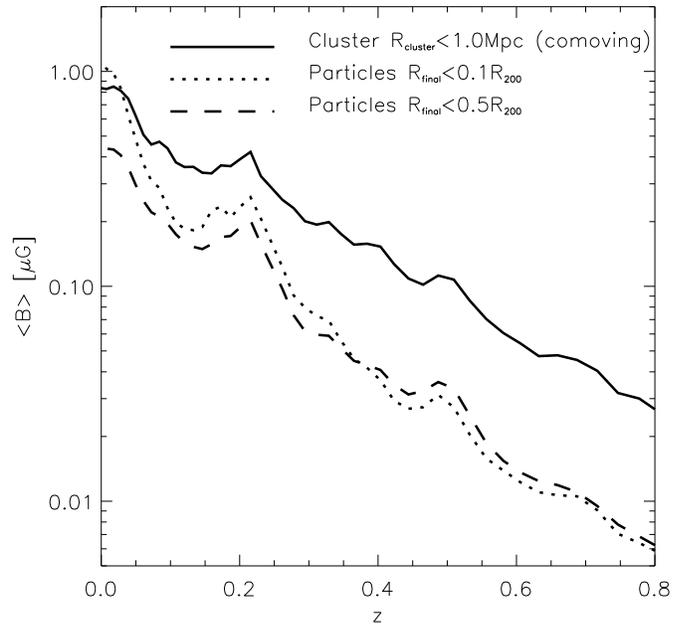}
\caption{The mean magnetic field in one cluster simulated in the SCDM
  cosmology ($h=0.5$) of those gas particles which at the end of the
  simulation populate spheres of ten and fifty per cent of the virial
  radius (dotted and dashed lines) around the cluster centre is shown
  as a function of redshift. The solid line is the mean magnetic field
  within a sphere of comoving radius 1~Mpc which remains centred on
  the most massive cluster component.}
\label{fig_evol_neu}
\end{figure}

\section{\label{sec:5} Magnetic field evolution}

This section describes how the magnetic fields evolve during the
formation of galaxy clusters in our simulations. The field evolution
is driven by structure formation; in particular, major mergers have a
strong influence on the how the field evolves.

\subsection{Magnetic field evolution}

We follow two different approaches for describing the evolution of the
magnetic field in our simulations. First, we look at each individual
given cluster and measure how the magnetic field strength in the
cluster (e.g.~within the core) evolves with time. On the other hand,
we can ask how the magnetic field evolves which is carried by the gas
that finally ends up in the cluster core. The first approach could be
considered as Eulerian, the second as Lagrangian.

Figure~\ref{fig_evol_neu} shows the results of both approaches for one
cluster simulated in the SCDM cosmology ($h=0.5)$). The solid line
shows the mean magnetic field within a sphere of 1~Mpc radius around
the cluster centre. More precisely, it shows the mean magnetic field
within a sphere of 1~Mpc comoving radius, centred on the most massive
halo contributing to the final cluster. The dotted and dashed lines show as
functions of time the mean magnetic fields carried by all particles
which finally fall into spheres around the cluster centre with radii
of ten and fifty per cent of the virial radius. It is clearly seen
that the magnetic field in the core of the evolving galaxy cluster is
stronger than the magnetic field in the cluster surroundings, which
are finally accreted by the galaxy cluster. All three curves show
pronounced maxima which coincide with major mergers at redshifts $0.5$
and $0.2$, and likewise at the second core passage of a major merger
at redshift $0.7$. Also, the combined effect of a second core passage
and a merger at redshift $\sim0.2$ can be seen here, and more
explicitly in Fig.~\ref{evol_biga}. The magnetic field in the cluster
cores and in related shocks during mergers is increased
substantially. Even the mean field within a large volume can be
increased by a factor of a few, as can be read off
Fig.~\ref{fig_evol_neu}.

\begin{table}[t]
\begin{center}
\begin{tabular}{|c|c|c|c|}
\hline
SCDM & $5$ & $20$ & $50$ \\
\hline
low B   & $-0.23$ & $-0.33$ & $-0.31$ \\
medium B& $-0.09$ & $-0.15$ & $-0.15$ \\
high B  & $-0.04$ & $-0.07$ & $-0.08$ \\
\hline
$\Lambda$CDM    &  $5$   &  $20$    & $50$ \\
\hline
low B   & $-0.29$ & $-0.34$ & $-0.42$ \\
medium B& $-0.16$ & $-0.18$ & $-0.24$ \\
high B  & $-0.09$ & $-0.11$ & $-0.15$ \\
\hline
\end{tabular}
\end{center}
\caption{This table summarises the slopes of the linear fits shown in
  Fig.~\ref{evol_eds} for different cosmologies and different
  Faraday-rotation thresholds.}
\label{tab_grow}
\end{table}

\subsection{Influence of mergers on observables}

We now investigate how the projected patterns of Faraday rotation
measures evolves. In doing so, we calculated synthetic rotation
measure maps for 60 epochs between redshifts $0.8$ and $0$
(today). For each epoch, we have one map for each of the three
different projection directions (chosen to be the coordinate axes). In
each map, we calculate the total area of those regions in which the
rotation measure exceeds a given threshold (in other words, we compute
the measure of the excursion set of the Faraday rotation).

\begin{figure}[t]
  \includegraphics[width=0.49\textwidth]{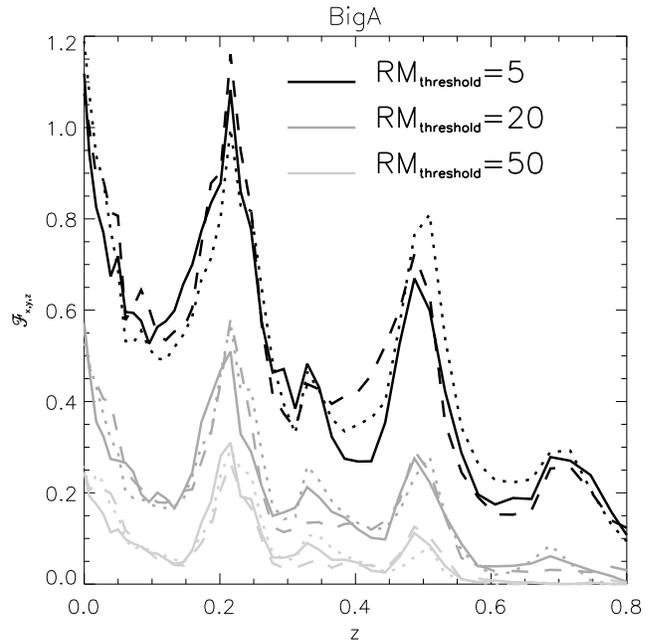}
\caption{This figure shows how the area, covered by a Faraday rotation
  measure exceeding thresholds of 5 (upper bundle), 20 (middle bundle)
  and 50~rad/m$^2$ (lower bundle) evolves with time. The area is
  normalised by the area emitting 95\% of the cluster X-ray
  luminosity.  The different line types distinguish the three
  different projection directions.} \label{evol_biga}
\end{figure}

Figure~\ref{evol_biga} shows this area as a function of time (or
redshift), normalised to the area which emits 95\% of the X-ray
luminosity. The different line types distinguish between the three
different projection directions, the three bundles of lines are for
rotation measure thresholds of 5, 20, and 50~rad/m$^2$. The general
tendency for this quantity to grow with time is evident, as is
expected from the growing magnetic field. The structure seen in the
time evolution is clearly related to merger events. All peaks on these
curves correspond to a core passage during a merger event. Likewise,
the second core passage after the merger causes the area to peak. The
signal does not strongly depend on the projection direction, and the
sequence of peaks is present for all rotation-measure thresholds
chosen. This provides clear evidence that merger events strongly
contribute to the amplification of the magnetic field, an effect which
is confirmed by other simulations (Roettiger et al. 1999). Since this
effect is visible in all projections, this also applies to mergers
along the line-of-sight, whose X-ray contours look quite regular. This
implies that determining the area covered by Faraday rotation measures
above a fixed threshold may help constraining cluster mergers in such
configurations.

\subsection{Evolution in different cosmologies}

\begin{figure*}[t]
  \includegraphics[angle=0,width=0.49\textwidth]{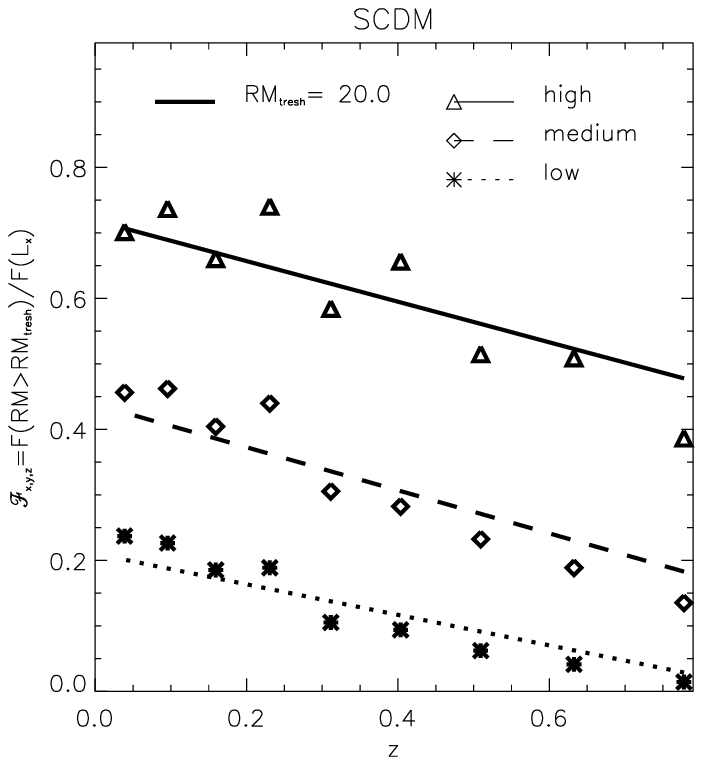}
  \includegraphics[angle=0,width=0.49\textwidth]{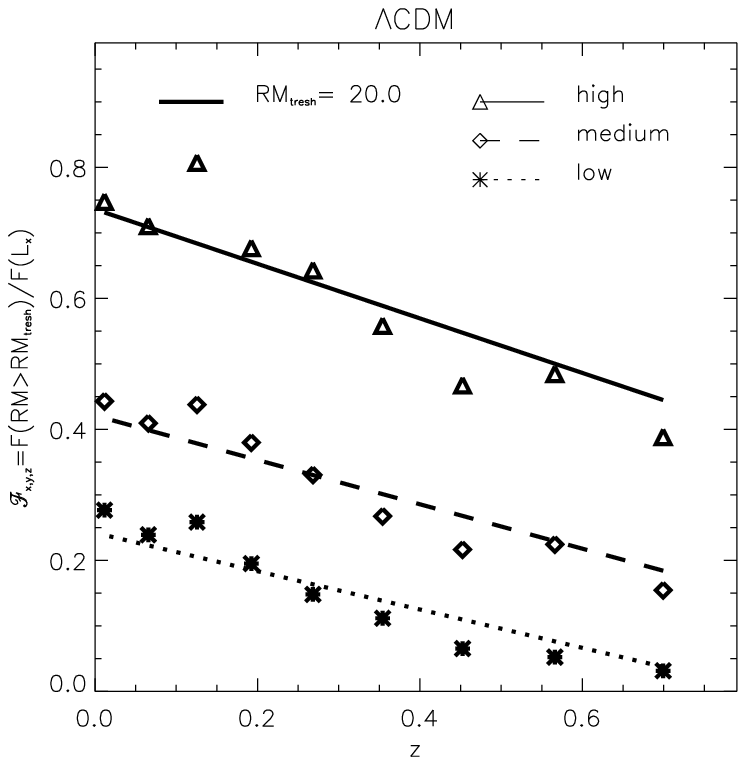}
\caption{This figure shows the same quantity as shown in
  Fig.~\ref{evol_biga}, but only for the one threshold of
  20~rad/m$^2$, and averaged over the three projection directions and
  over the ten clusters per sample. The different line types represent
  the different sets of models for the initial magnetic field. The
  left panel shows results for the SCDM cosmology, the right panel for
  the $\Lambda$CDM cosmology.} \label{evol_eds}
\end{figure*}

Figure~\ref{evol_eds} shows the same area, but averaged over all ten
realizations. The sequence of peaks is smoothed out, leaving the
general trend. The different initial field configurations are
indicated by different types of symbol, as indicated in the plot. We
overlay fits showing that the growth is very well fit by linear
functions. Table~\ref{tab_grow} summarises the gradient inferred from
these fits for different cosmologies, different initial field
configurations and different thresholds.

\section{\label{sec:6} Comparison with measurements}

\begin{figure*}[th]
 \includegraphics[angle=0,width=0.49\textwidth]{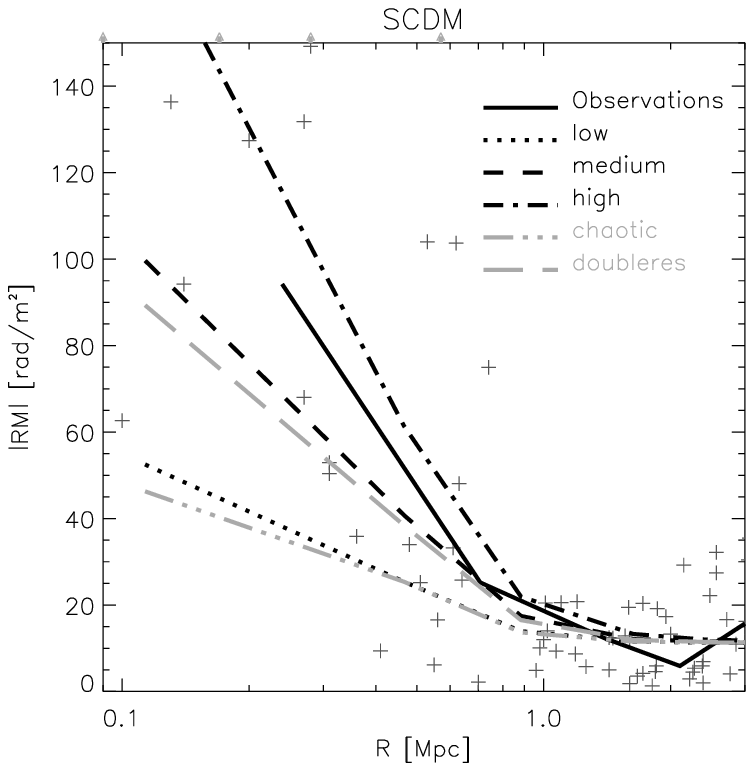}
 \includegraphics[angle=0,width=0.49\textwidth]{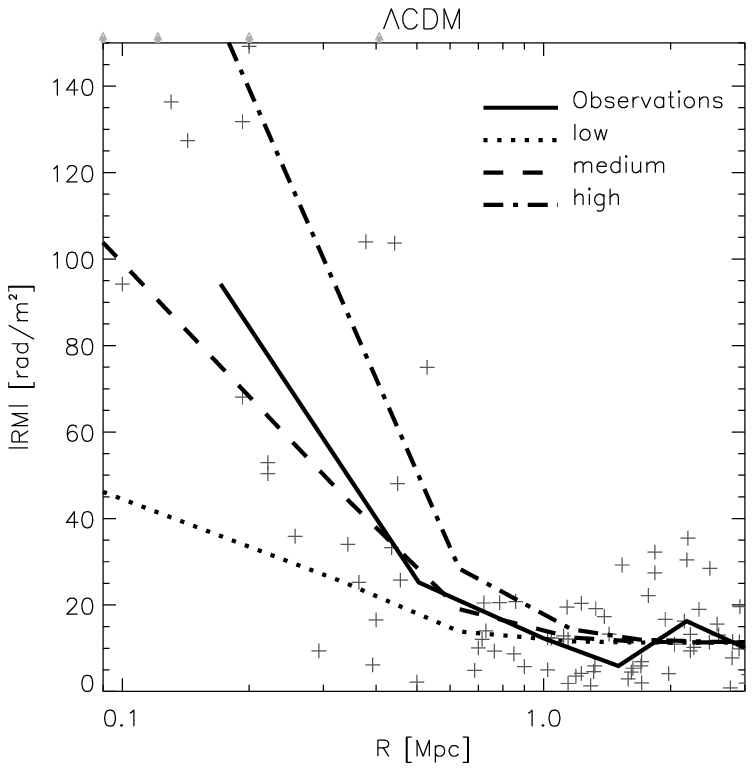}
\caption{This figure shows a comparison of observed Faraday rotation
  measures with our simulations. Plotted is the absolute value of the
  observed rotation measures as a function of the distance to the next
  Abell cluster (crosses), taken from Clarke et al.~(2001). The solid
  line marks the sample median of the observations. The superposed
  lines are calculated from synthetic Faraday rotation measurements
  taken from our simulations in an attempt to mimic the
  observations. The different line types distinguish between the
  different initial magnetic field configurations used in the
  simulations. The left and right panels show SCDM and $\Lambda$CDM
  results, respectively. The Medians were calculated in bins
  containing 15 data points each. The shape of the synthetic measurements in
  both cosmologies follow very well the observations. To perfectly
  match the observations in this comparison, a magnetic field
  somewhere in between the ``{\em medium\/}'' and the ``{\em high\/}''
  models is needed.}
\label{comp_all}
\end{figure*}

Figure~\ref{comp_all} compares the measurements of Clarke et
al.~(2001) with the synthetic Faraday rotation measures taken from our
simulations. These new measurements are in excellent agreement with
previous measurements collected by Kim et al.~(1990), and are also
compatible with Faraday-rotation measurements in individual clusters
like Abell~119 by Feretti et al.~(1999).

To calculate the synthetic signal, we computed three rotation-measure
maps for the three spatial projects for each cluster on a grid of
$60^2$ cells, covering $5\times5\,$Mpc (physical units). For each
cosmology and all initial magnetic field configurations, the
Faraday-rotation measures are sorted in radial bins and combined over
all ten realizations. The simulated values compare very well with the
observations. We calculated the signal from the observations between 3
Mpc and 10 Mpc as a reference (11~rad/m$^2$) and added this as a
background to the synthetic data. The synthetic curves very well
follow the shape of the measurement distributions. It is also easily
seen that the initial magnetic field strength required to reproduce
the observations falls somewhere between the ``medium'' and the
``high'' models.

Again, it is obvious that the difference for the two different initial
field configurations, namely homogeneous and chaotic, is negligible.
Also, the two different resolutions lead to statistically the same
signal, but as the mass resolution is only changed by a factor of two,
this does not resemble a convergence study.

\section{Summary and conclusions}

We performed cosmological, magneto-hydrodynamic simulations of galaxy
clusters to study the evolution and final structure of magnetic fields
frozen into the intracluster medium. In total, we investigated 100
cluster models of different masses in two different cosmologies
(standard and open CDM), varying the structure and the strength of the
initial magnetic field configuration. Our results can be summarised as
follows:

\begin{itemize}

\item Starting with magnetic fields of order $10^{-9}\,\mathrm{G}$ at
  an initial redshift of $z_\mathrm{ini}=15$ (20 for $\Lambda$CDM),
  final fields reach micro-Gauss strength in cluster cores at redshift
  $z=0$. This amplification factor of $\sim10^3$ was found and
  discussed earlier (Dolag et al. 1999). It exceeds the amplification
  factor expected from spherical collapse by an order of magnitude,
  which can be traced back to field amplification in shear flows.

\item Radial profiles of the magnetic field strength closely follow
  the cluster density profiles at radii of $\sim200\kpc$ and
  beyond. At smaller radii, the magnetic field strength profiles
  flatten. This reflects the behaviour of the gas distribution, which
  also develops a flat core despite the dark-matter density cusp
  because of the finite gas pressure. In a purely spherical collapse,
  the magnetic field strength is expected to scale with density raised
  to the $2/3$ power, flattening the profile. Additionally, the
  amplification by shear flows works more efficiently when the field
  strengths are small and therefore the fields get more relative
  amplification in the outer parts, leading to further flattening of
  the magnetic field profile compared to the gas profile. This happens
  independently of the strength and the structure of the initial
  magnetic seed field, and for both cosmologies investigated.

\item The correlation lengths of the magnetic fields in cluster cores
  are of order $\sim50\kpc$. Similarly, the length scale of field
  reversals along lines-of-sight through cluster cores is of order
  $\sim100\kpc$. These length scales show no significant dependence on
  the initial field strength or configuration, or on cosmology.

\item The power spectrum of the magnetic field can closely be
  approximated by a power law, $P_\mathrm{B}(k)\propto k^{-n}$, with a
  power-law exponent of $n\sim-2.7$. This is steeper than expected for
  Kolmogorov or MHD turbulence, but close to the expectation for
  two-dimensional Navier-Stokes turbulence. A possible explanation is
  the occurrence of essentially two-dimensional shear flows and their
  importance for the magnetic field evolution in clusters. Although we
  do not expect this to be a strong effect, the high numerical
  viscosity in SPH codes also tends to steepen the power spectrum.

\item Predominantly driven by merger events, the mean magnetic field
  in cluster cores grows roughly exponentially with redshift. Between
  redshifts $z=0.8$ and $z=0$, the growth is well approximated by
\begin{equation}
  |\vec B|(z)\approx10^{-2.5\,z}\,\muG\;.
\end{equation}
  Superposed on this general trend is the effect of mergers, during
  which the magnetic field strength goes through pronounced maxima.

\item Field amplification by mergers shows up prominently in
  observable quantities like the Faraday rotation. During mergers, the
  projected cluster area in which substantial Faraday rotation occurs
  rises sharply, and it drops during the subsequent relaxation phases.

\item Simulated Faraday rotation measures show that all our cluster
  models are in excellent agreement with the largest available sample
  of observed Faraday-rotation measurements, although the samples are
  still too small to provide narrow constraints on, e.g., the initial
  strength of the magnetic seed fields.

\end{itemize}

Our results therefore show that the strength of intracluster magnetic
fields grows rapidly with redshift, and it closely follows the cluster
density profiles outside a core region of $\sim200\kpc$ radius. The
structure of the field can be characterised by an auto-correlation or
reversal scale of order $50-100\kpc$, well above the spatial
resolution of our numerical simulations. The power spectrum of the
magnetic field falls steeply, but the distribution of energy density
with respect to $k$ scales as $k^3\,P(k)$ and thus remains
monotonically increasing. This indicates that most of the magnetic
energy density is located on small scales.

The independence of these results on cosmology and initial field
structure demonstrates that the final intracluster fields are entirely
dominated by the cluster collapse, thus confirming and extending the
previous findings (Dolag et al. 1999). The importance of mergers for
observable signatures of intracluster magnetic fields, and their steep
evolution with redshift, suggest important observational tests for the
scenario underlying the formation and evolution of our simulated
galaxy clusters.

\end{document}